# Topography, climate, land cover, and biodiversity: Explaining endemic richness and management implications on a Mediterranean island

**Aristides Moustakas[1,*] Ioannis N Vogiatzakis[2]**

1. Natural History Museum of Crete, University of Crete, Heraklion, Greece

2. School of Pure & Applied Sciences, Open University of Cyprus, PO 12794, Nicosia, Cyprus

* Corresponding author:

Aristides (Aris) Moustakas

Email: arismoustakas@uoc.gr

**Abstract**

Island endemism is shaped by complex interactions among environmental, ecological, and evolutionary factors, yet the relative contributions of topography, climate, and land cover remain incompletely quantified. We investigated the drivers of endemic plant richness across Crete, a Mediterranean biodiversity hotspot, using spatially explicit data on species distributions, topographic complexity, climatic variability, land cover, and soil characteristics. Artificial Neural Network models, a machine learning tool, were employed to assess the relative importance of these predictors and to identify hotspots of endemism. We found that total species richness, elevation range, and climatic variability were the strongest predictors of endemic richness, reflecting the role of biodiversity, topographic heterogeneity, and climatic gradients in generating diverse habitats and micro-refugia that promote speciation and buffer extinction risk. Endemic hotspots only partially overlapped with areas of high total species richness, indicating that total species richness was the optimal from the ones examined, yet an imperfect surrogate. These environmentally heterogeneous areas also provide critical ecosystem services, including soil stabilization, pollination, and cultural value, which are increasingly threatened by tourism, renewable energy development, land-use change, and climate impacts. Our findings underscore the importance of prioritizing mountainous and climatically variable regions in conservation planning, integrating ecosystem service considerations, and accounting for within-island spatial heterogeneity. By explicitly linking the environmental drivers of endemism to both biodiversity patterns and ecosystem function, this study provides a framework for evidence-based conservation planning in Crete and other Mediterranean islands with similar geological and biogeographic contexts.

**Introduction**

Although islands cover only a small fraction of the Earth's surface, they harbor a disproportionate share of biodiversity, with over one-third of global hotspots located on islands (Myers et al., 2000; Schrader et al., 2024). Their restricted area, isolation, and human pressures make them highly vulnerable to global change (IPBES, 2019; WWF, 2022; Vogiatzakis et al., 2023). Yet the responses of island biodiversity to land use and climate change remain poorly understood (Moustakas et al., 2025), hindering effective policy translation (Cámara-Leret & Dennehy, 2019). Endemic species, confined to specific regions, are of particular concern due to their unique evolutionary histories (Cañadas et al., 2014).

Climate strongly influences species evolution and distribution (Parmesan, 1996; Lawlor et al., 2024). Species occurrences depend on temperature, precipitation, and climatic variability (Gutiérrez-Hernández & García, 2021; Gallou et al., 2023; Fonteyn et al., 2025). Climatic heterogeneity can act as a refugium, providing microclimatic buffering (Gavin et al., 2014; Gallou et al., 2023). However, climate-only models often perform poorly (Lennon, 2000; Beale & Lennon, 2012), as species responses depend on interactions with other abiotic and biotic factors (Lawlor et al., 2024). Broader analyses are therefore needed (Peng et al., 2024).

Land cover and habitat also determine endemic distributions (Chauvier et al., 2021; Gábor et al., 2024). While endemics are usually linked to natural habitats, some persist in agricultural or human-modified environments (Ondoño et al., 2015; Adu-Acheampong et al., 2016). Habitat heterogeneity can enhance diversity but may disadvantage habitat specialists (Matthews et al., 2014). Topography shapes distributions by offering refugia in steep or inaccessible areas (Trigas et al., 2013; Irl et al., 2015; Fitzsimons & Michael, 2017). Elevation and soil further influence plant endemism (Bachman et al., 2004; Bárcenas-Argüello et al., 2013; Hulshof & Spasojevic, 2020; Chrysostomou et al., 2024).

Patterns of species richness in islands have been long studied (MacArthur & Wilson, 1967; Benavides Rios *et al.*, 2024). Island diversification has been linked to overall species richness, supporting the notion that "biodiversity begets biodiversity" (Emerson & Kolm, 2005; van Holstein & Foley, 2024). Endemic richness often correlates with total richness because both depend on island attributes such as area, climate, and habitat (Cadena et al., 2005; Schluter & Pennell, 2017; Beierkuhnlein, 2024).

Island endemics are not only markers of unique evolutionary processes but also key providers of ecosystem services, including pollination, soil stabilization, water regulation, and nutrient cycling, as well as cultural and economic benefits. Their restricted ranges make them especially vulnerable, meaning that even small-scale disturbances can have disproportionate impacts. The loss of endemics threatens both ecosystem integrity and human well-being, increasing risks under global change.

Recognizing their dual ecological and societal roles highlights the need to incorporate them into conservation planning and ecosystem service assessments. These priorities align with international frameworks such as the IPBES global assessments (IPBES, 2019) and the EU Biodiversity Strategy for 2030 (European Commission 2020), which emphasize the protection and restoration of species-rich habitats. Safeguarding endemic-rich areas through protected networks, sustainable land use, and climate adaptation is therefore critical for meeting biodiversity targets and maintaining the ecosystem services that support human well-being (IPBES, 2019).

Understanding endemic distributions is complex, as interacting drivers often violate statistical assumptions (Calude & Longo, 2017). Computational approaches that manage multiple, correlated variables provide plausible alternatives to these issues (Kar & Dwivedi, 2020). The growing availability of spatial data and environmental sensors creates new opportunities (Moustakas, 2017; Karimi, 2024), while interdisciplinary, data-driven methods promise deeper insights into species–environment relationships (Moustakas & Katsanevakis, 2018; Leonelli & Tempini, 2021).

We use the island of Crete as a case study because of its exceptional plant endemism, diverse land cover, complex topography, and pronounced anthropogenic pressures (Vogiatzakis et al. 2000; Papanastasis & Kazaklis, 1998). Combining a spatially explicit atlas of Cretan flora with environmental predictors, we applied machine learning to disentangle the independent effects of multiple, potentially correlated drivers. Specifically, we address four questions: (1) How do biophysical parameters, including climate, land cover, topography, and soil, shape endemic species richness? (2) To what extent does total species richness predict endemic richness? (3) How are endemic species distributions clustered with respect to environmental variables and overall biodiversity? and (4) What are the implications of these patterns for ecosystem services and biodiversity management? We hypothesize that endemic richness is best explained by the joint influence of climate, land cover, topography, and soils, rather than climate alone, and that areas of high overall richness also support more endemic species. By testing this hypothesis, we aim to generate insights that can guide both ecological theory and applied conservation policy.

## 2. Methods

### 2.1 Study Area & data overview

The study focused on Crete and its surrounding islets ('Crete'), with the island centre located at coordinates 35.2401° N, 24.8093° E. Crete is a Mediterranean biodiversity hotspot and one of the ten red alert areas for conservation in the region (Médail & Diadema, 2009). Endemic plant species account for over 10% of the native vascular flora, with approximately (depending on the study and plant taxonomy) 1,700 vascular plant species documented across the island's 8,374 km² (Turland *et al.*, 1993; Lazarina *et al.*, 2019; Menteli *et al.*, 2019). Crete's topography is dominated by three mountain massifs: Psiloritis, the highest at 2,456 m, and Lefka Ori (White Mountains), the largest, with 15 peaks exceeding 2,200 m (Vogiatzakis et al., 2003). Elevation varies steeply, particularly in the south-west, creating strong environmental gradients (Vogiatzakis et al., 2003). Mean annual precipitation (mean = 732 mm) generally increases toward the west and north and peaks at higher elevations, whereas temperature (mean = 16.69 C) rises southward and is highest at lower elevations. Coastal areas are heavily modified by urban and tourist infrastructure, while the majority of the human population resides in northern urban centres. Agriculture is widespread, predominantly olive groves and vineyards. Crete's soils mainly derive from its carbonate geology, with shallow stony Leptosols dominating the island. Upland areas have thin, rocky soils, while terra rossa occurs on limestone plateaus. Deeper alluvial soils in valleys and plains support agriculture, and small serpentine outcrops host distinctive endemic-rich vegetation. (Yassoglou *et al.*, 2017).

### 2.1.2 Plant Distribution Data

We used the spatially explicit Flora of the Cretan Area dataset, which divides Crete into 162 grid cells of 8.25 × 8.25 km (Turland et al., 1993; Chilton & Turland, 2004, 2008). These cells do not distinguish between land and water though in the vast majority contains land exclusively. Cell adjustments were made as follows: (1) coastal cells with <10% land area were merged with adjacent cells; (2) very small islands and islets were incorporated into neighbouring coastal cells; (3) isolated islands were included in modified squares so that each island fell entirely within a single cell (Turland et al., 1993). The final cells has an approximately equal land surface area (differences < 2%). of All spatial data were harmonized to an 8.25 km resolution.

*2.1.3 Species Richness*

Plant species richness per cell was calculated from the Flora of the Cretan Area (Turland et al., 1993; Chilton & Turland, 2004, 2008). Total species richness included 1,706 species. Total species richness was computed per cell as the sum of all species present in the cell. Endemic species richness included 174 species. Endemic species richness was computed per cell as the sum of all endemic species present in the cell.

*2.1.4 Climate*

Climatic variables were obtained from WorldClim (Hijmans et al., 2009) and aggregated from 1 km to 8.25 km by averaging all cells within each grid. Variables included annual mean temperature, mean temperature of warmest and coldest quarters, temperature range, annual mean precipitation, precipitation of wettest and driest quarters, and precipitation range. Data from 2010 were used to approximate the temporal context of the 2008 flora records.

*2.1.5 Land Cover*

CORINE Land Cover Level 3 (EEA, 2000) at 100 m resolution was used to quantify land cover composition per cell via Patch Analyst in ArcGIS. We calculated richness and percentage cover for natural, agricultural, and artificial land cover types, as well as total land cover richness. The 2012 snapshot was used to align temporally with the species dataset.

*2.1.6 Topography*

Mean elevation and elevation range per cell were computed using 100 m elevation intervals. Mean elevation was the average of all points per cell; elevation range was the difference between maximum and minimum values.

*2.1.7 Soil*

Soil data were sourced from SoilGrids (Hengl et al., 2014) at 1 km resolution and aggregated to 8.25 km. For each cell, the dominant soil type and soil richness (number of unique soil types) were recorded.

All variables together with their mean values and units used as independent explanatory covariates in the analysis are listed in Table 1.

**Table 1.** Description and mean value of independent explanatory variables of endemic species richness deployed in the analysis.

| Nr | Explanatory variables | Mean value | Abbreviation | Unit |
|---|---|---|---|---|
|  | **Climate** |  |  |  |
| 1 | Mean temperature | 16.6901 | Temp_Mean | C |
| 2 | Temperature range | 13.0574 | Temp_Rang | C |
| 3 | Temperature coldest quarter | 10.6043 | Temp_Cold | C |
| 4 | Temperature warmest quarter | 23.6617 | Temp_Warm | C |
| 5 | Mean precipitation | 732.111 | Precip_Mean | mm |
| 6 | Precipitation range | 379.056 | Precip_Range | mm |
| 7 | Precipitation wettest quarter | 389.5 | Precip_Wet | mm |
| 8 | Precipitation driest quarter | 10.4444 | Precip_Dry | mm |
|  | **Land cover** |  |  |  |
| 9 | Natural land cover types richness | 4.20988 | Natural_Div | Num |
| 10 | Agricultural land cover types richness | 3.34568 | Agric_Div | Num |
| 11 | Artificial land cover types richness | 0.876543 | Artif_Div | Num |
| 12 | Total land cover types richness | 8.4321 | Hab_Div | Num |
| 13 | Natural land cover percentage | 58.8474 | Natural_Cover | % |
| 14 | Agricultural land cover percentage | 39.231 | Agric_Cover | % |
| 15 | Artificial land cover percentage | 1.742 | Artif_Cover | % |
|  | **Soil** |  |  |  |
| 16 | Predominant soil type |  |  | Factor |
| 17 | Soil types richness | 3.45679 | Soil_Div | Num |
|  | **Topography** |  |  |  |
| 18 | Mean elevation | 296.302 | Alt | m |
| 19 | Elevation range | 736.969 | Alt_Range | m |
|  | **Biodiversity** |  |  |  |
| 20 | Total species richness | 122.525 | Total_Spp | Num |

*2.2 Analysis*

2.2.1 Principal Components Analysis (PCA)

To explore patterns of environmental and habitat variation across study sites, we conducted a Principal Component Analysis (PCA) on all explanatory variables, except soil type that is a factor (Demšar *et al.*, 2013). All variables were standardized to zero mean and unit variance to ensure comparability. PCA was performed to identify potential multicollinearity and summarize major gradients of variation among variables and sites. The first four principal components (PC1–PC4) were examined, with variable loadings >0.3 considered influential, and were used to visualize site clustering and explore related ecological patterns (Demšar *et al.*, 2013).

2.2.2 Artificial Neural Network Model

Predictor variables were standardized (mean = 0, standard deviation = 1) to improve model convergence. Categorical (factor) variables were transformed into binary indicator variables. We employed feedforward artificial neural networks (ANNs) regression to model endemic species richness as a function of the predictor variables (Hasson *et al.*, 2020). The ANN architecture consisted of an input layer corresponding to the number of predictor variables, one hidden layer with four neurons each, and an output layer with a single neuron representing predicted species richness (Ibnu Choldun R. *et al.*, 2020). Rectified Linear Unit (ReLU) activation functions were used for hidden layers, and a linear activation function was applied at the output layer to accommodate continuous response values (Razavi, 2021). All analyses were conducted in Python 3.11 using the TensorFlow (v2.x) and Keras libraries for ANN implementation.

2.2.3 Model Training and Validation

The dataset was randomly split into training (80%) and testing (20%) subsets. Model parameters were optimized using the Adam optimizer with a learning rate of 0.001 and Mean Squared Error (MSE) as the loss function. Early stopping based on validation loss was implemented, with a patience of 10–20 epochs and a minimum improvement threshold of 0.001 to prevent overfitting (Razavi, 2021). Hyperparameter tuning, including the number of hidden layers, neurons per layer, and regularization strength, was conducted using a grid search approach with five-fold cross-validation on the training data (Razavi, 2021).

*2.2.4 Model Evaluation*

Model performance was evaluated using the testing dataset with multiple metrics, including root mean squared error (RMSE), mean absolute error (MAE), and coefficient of determination ($R^2$). Variable importance was quantified using permutation importance, which measures the decrease in model performance when the values of a predictor are randomly permuted. This approach allowed us to identify the environmental variables that most strongly influenced predictions of endemic species richness in the neural network (Altmann *et al.*, 2010). To further investigate model performance, ANN model outputs were plotted versus endemic species richness data and a linear regression between ANN outputs and data and 95% confidence intervals were calculated and plotted. Residuals of ANN outputs were plotted versus endemic species richness data to quantify whether residuals are potentially biased towards lower or larger values. A linear regression between ANN residuals and data and 95% confidence intervals were calculated and plotted.

*2.3 Explaining machine learning outputs*

To examine spatial patterns and environmental drivers of endemic species richness, we plotted endemic species richness, and their top 20% explanatory variables as derived from machine learning outputs. Spatial overlap and distribution patterns were visually assessed using maps. Relationships between endemic species richness and their top 20% explanatory variables were quantified using Pearson correlation coefficients. Correlation strengths were interpreted to evaluate the degree to which each variable explained variation in endemic species richness across sites.

**3. Results**

*3.1 Principal Components Analysis (PCA)*

A principal component analysis (PCA) summarized explanatory variables across spatial locations (cells). The first two principal components captured distinct ecological gradients. PC1 represented a climatic gradient, with positive loadings for precipitation and elevation and negative loadings for temperature, distinguishing cool, wet highlands from warm, dry lowlands (Fig. 1a; 1b). PC2 contrasted human-modified versus natural landscapes, positively with high agricultural cover and diversity and negatively with natural cover (Fig. 1a; 1b). PC3 reflected habitat heterogeneity and species richness, where artificial habitats were associated with lower biodiversity. PC4 captured soil diversity and species richness, independent of elevation (Fig. 1a; 1b). These PCs provide a concise framework for understanding ecological gradients across the landscape.

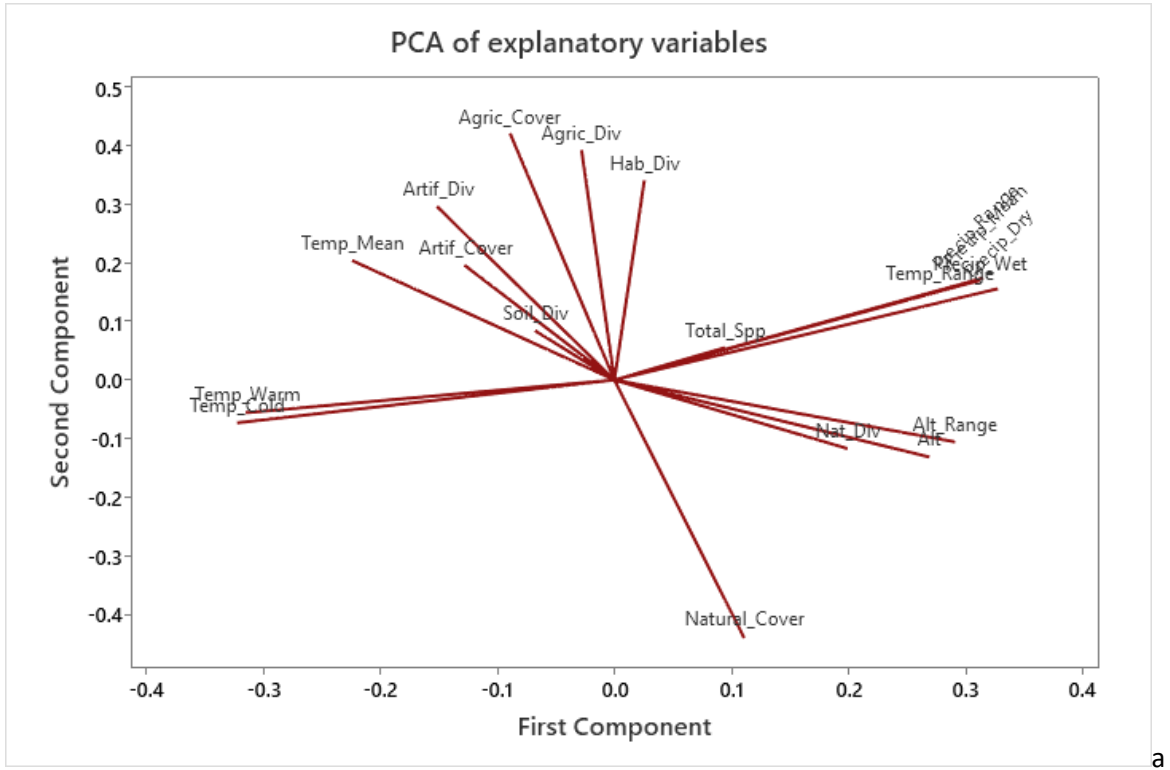
a

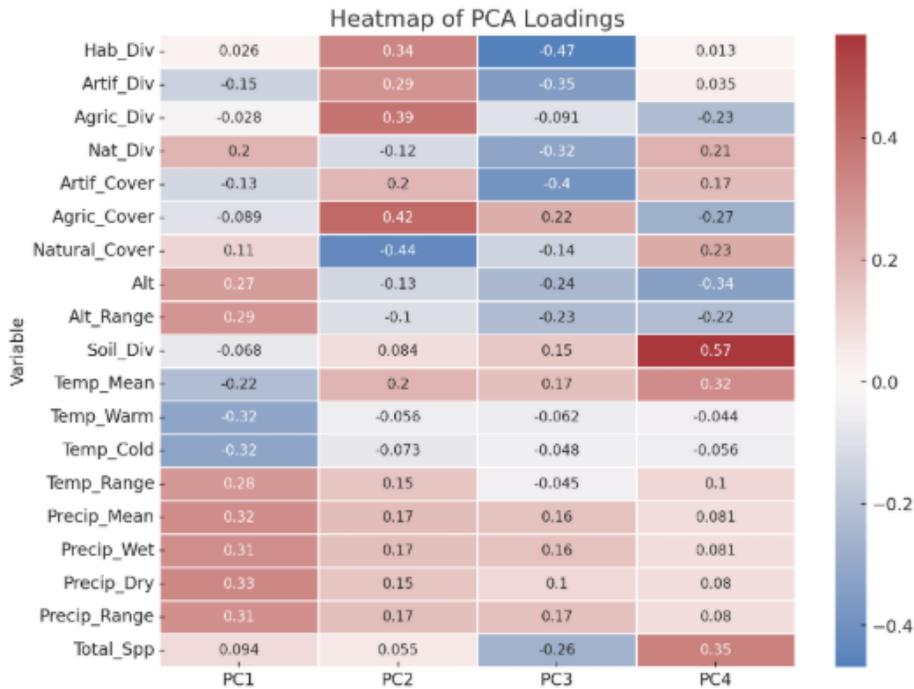

b

**Figure 1.** Principal Components Analysis (PCA) of explanatory variables. **a.** PCA biplot of PC1 and PC2 showing site scores and variable loadings; arrow length indicates the strength of each variable's influence. **b.** Heatmap of standardized environmental, habitat, and species richness variables.

*3.2 Machine learning ANN outputs*

Machine learning ANN model outputs indicated that the highest ranked variable for endemic species richness was total species richness, while elevation range, temperature range, elevation, and temperature during the coldest quarter followed (Fig. 2a). Each of these five variables covers up to 20% of normalised importance (Fig. 2a). Land cover diversity, soil diversity, temperature during the warmest quarter, and natural land cover percentage exhibited a normalised importance each up to 15% (Fig. 2a). Precipitation during the driest quarter, land cover type richness, and mean precipitation showed a normalised importance of up to 10% each (Fig. 2a). Mean temperature, agricultural land cover diversity, artificial land cover diversity, precipitation during the wettest quarter and precipitation range were the variables with the lowest normalised importance (Fig. 2a). ANN model fit exhibited a good fit in both the training and test data partitions (Table 2). ANN outputs were consistent with data values of endemic species richness across the range of data values (Fig. 2b) and model residuals were well behaved at all times (Fig. 2c).

**Table 2.** ANN model summary. For model validation data is divided into training and test partitions: the training set reveals dependencies, while the test set verifies model accuracy on new data not used for training.

| Data partition | Accuracy | Score |
|---|---|---|
| Training | MAE | 9.354 |
| | RMSE | 0.143 |
| | $R^2$ | 0.982 |
| Testing | MAE | 1.382 |
| | RMSE | 1.731 |
| | $R^2$ | 0.973 |

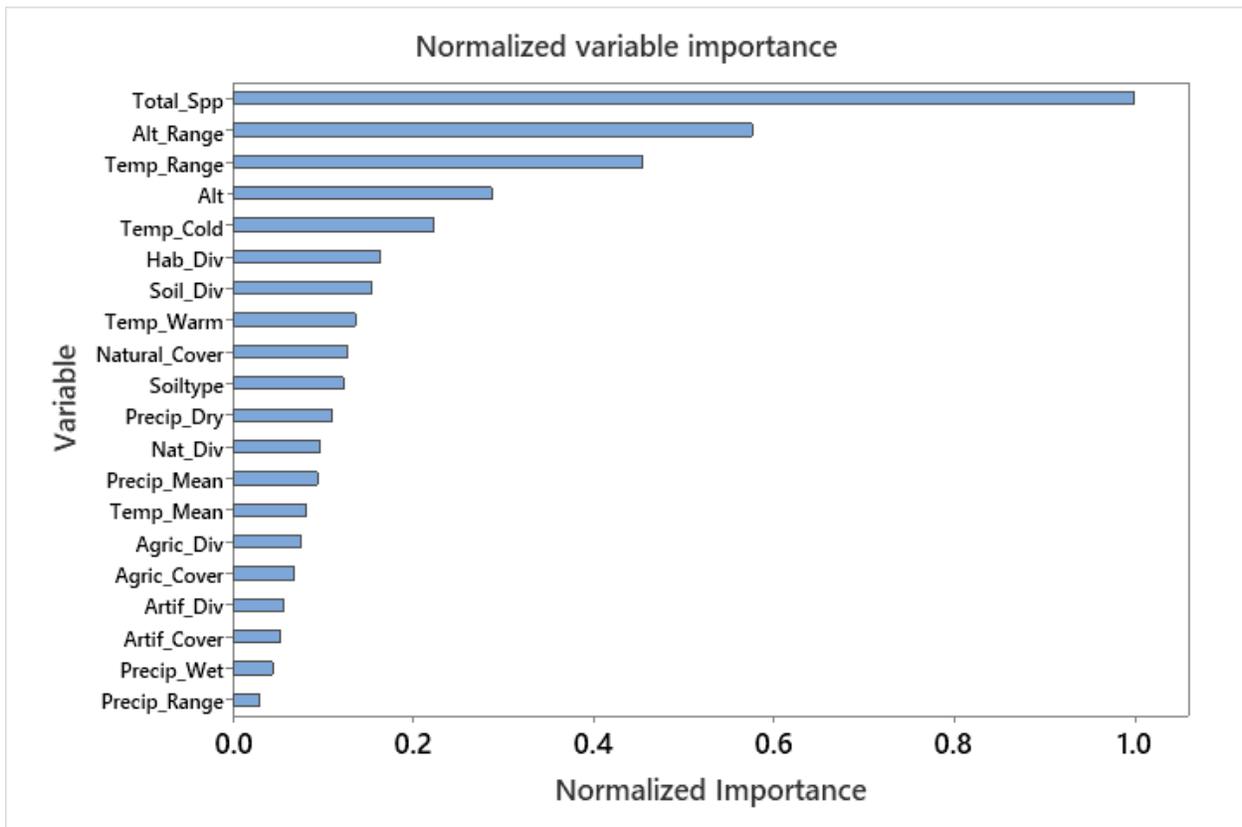

a

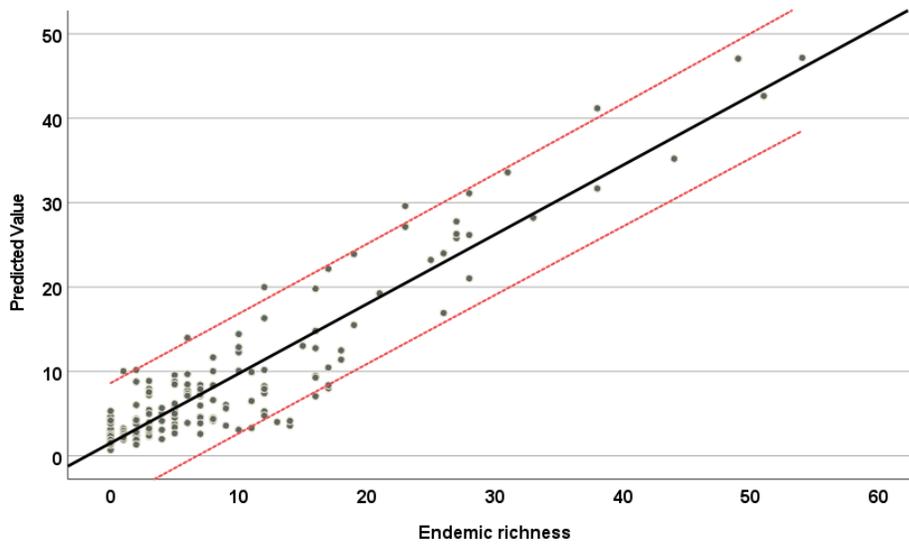

b

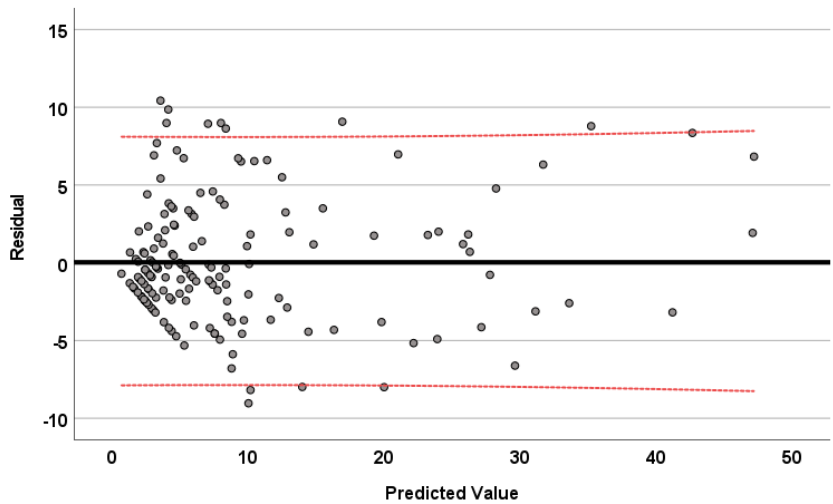

c

**Figure 2.** Outputs of Artificial Neural Networks (ANN) machine learning analysis. **a.** Relative variable importance chart measuring each variable's importance score standardised to the most important predictor. **b.** ANN model outputs (vertical axis) versus endemic species richness data (horizontal axis). Solid black line indicates a linear regression fit between ANN outputs and data while upper and lower dotted red lines indicate a 95% confidence interval. **c.** ANN model residuals (vertical axis) versus ANN outputs. Solid black line indicates a linear regression fit between ANN outputs and data while upper and lower dotted red lines indicate a 95% confidence interval.

*3.3 Explaining machine learning outputs*

*3.3 Explaining machine learning outputs*

The spatial distribution of endemic species richness shows a similar pattern to that of total species richness, with many locations near or below the mean and fewer sites with high richness (Fig. 3a). However, locations with high endemic richness generally do not coincide with areas of high total species richness (Fig. 3a). The distribution of elevation range differs from that of absolute elevation (Fig. 3a); most locations fall below 1200 m in elevation range, but some sites exhibit very high ranges, reflecting areas that span from sea level to mountain peaks (Fig. 3a). Spatially, locations with high elevation range often overlap with sites of high elevation (Fig. 3a).

Endemic species richness has a positive correlation with total species richness, elevation and elevation range with correlation scores over 50% (Fig. 3b). This indicated a large increase of endemic species richness per unit of increase in each of those three explanatory variables (Fig. 3b). There is also a positive correlation between endemic species richness and temperature ranges with a more moderate score (Fig. 3b) Endemic species richness is negatively correlated with temperature during the coldest quarter, with correlation coefficient of -44% (Fig. 3b).

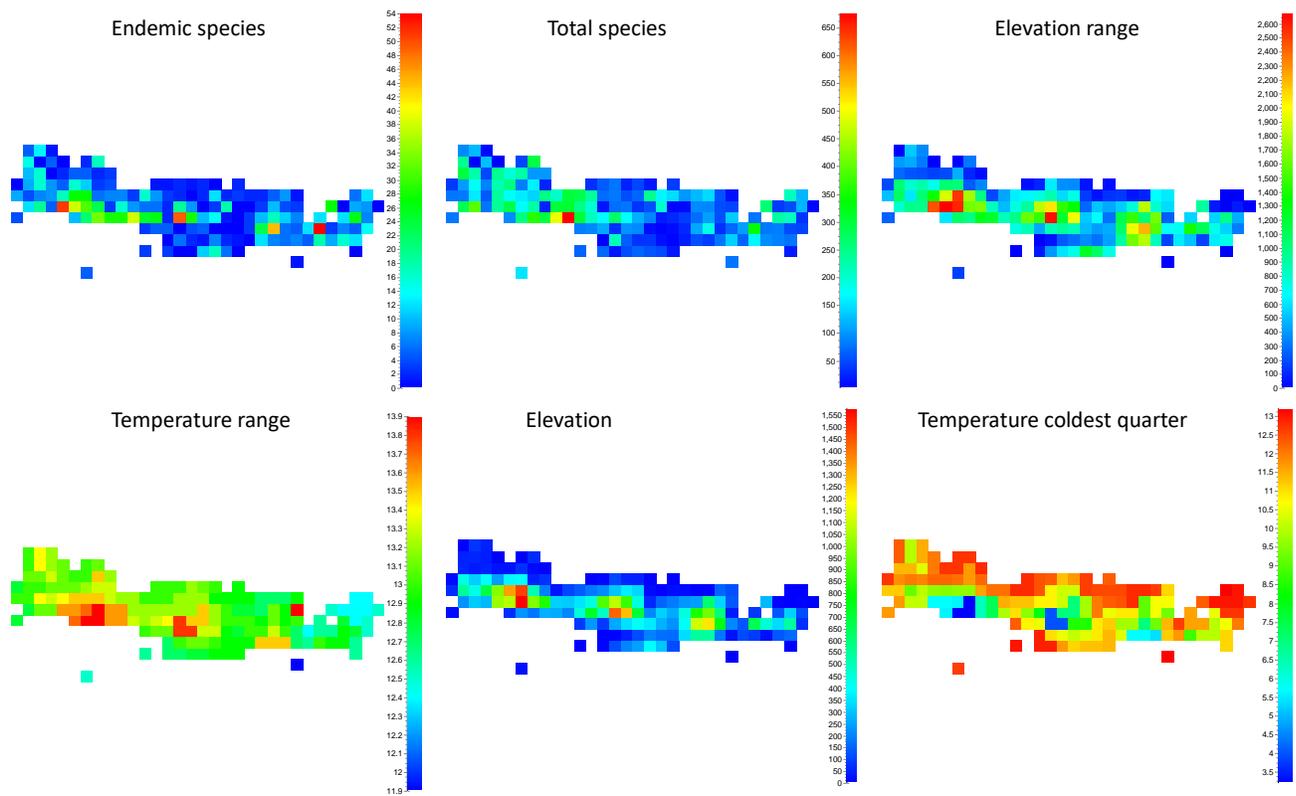

a

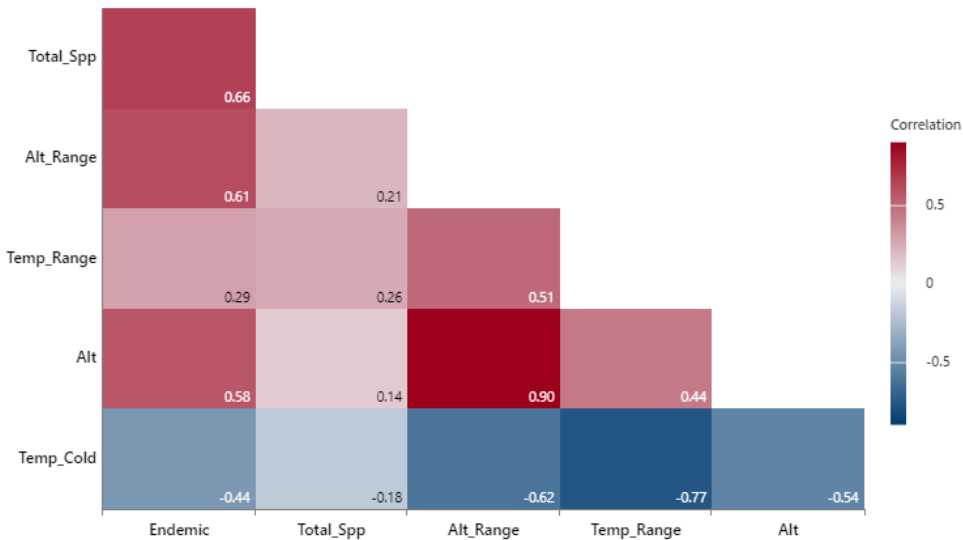

b

**Figure 3.** a. Maps of endemic species richness and the top 20% of explanatory variables as quantified by machine learning. These variables are: Total species richness, Elevation range, Temperature range, Elevation, and Temperature of the coldest quarter. **b.** Correlograms between Endemics species richness and the top 20% of explanatory variables as quantified by machine learning

## 4. Discussion

Endemic richness in Crete is strongly shaped by environmental heterogeneity, with topographic complexity and climatic variability emerging as the dominant predictors together with total species richness (Testolin *et al.*, 2024). By applying neural network models to integrate multiple environmental layers across the island, we were able to disentangle the relative contributions of topography, climate, land cover, and total species to endemic diversity (Irl et al. 2015). This methodological approach is particularly insightful because it captures non-linear relationships and interactions among predictors that traditional statistical techniques often overlook, providing a more nuanced understanding of the processes structuring biodiversity (Kar & Dwivedi, 2020). The robustness of the models, despite modest sample sizes, demonstrates the value of machine learning for identifying patterns in complex ecological systems in a wide range of questions (Alotaibi & Nassif, 2024; Moustakas et al. 2025). These results reinforce the view that diverse niches and refugial conditions drive island endemism, while also showing that hotspots of endemics and total species do not fully overlap (Harrison & Noss, 2017). Thus, endemic diversity reflects distinct ecological and evolutionary processes, underscoring the need for conservation strategies that go beyond protecting areas of high overall species richness (Jaén Molina *et al.*, 2025).

Moreover, the ability of these models to highlight areas where endemism diverges from overall richness offers a powerful tool for evidence-based management, ensuring that conservation priorities are informed by the ecological processes most critical to long-term persistence.

### 4.1 Explaining endemic species patterns

*4.1.1 Biodiversity*

Endemic richness increases with total species richness, consistent with the idea that "biodiversity begets biodiversity" (Emerson & Kolm, 2005; van Holstein & Foley, 2024). Because factors promoting overall richness often also benefit endemics, treating endemic and non-endemic species as functionally equivalent may be reasonable in island systems (Cutts et al., 2023). However, endemic hotspots only partly coincide with overall richness (Kougioumoutzis et al., 2021), as different drivers shape richness and turnover (Lazarina et al., 2019). Invasive richness also rises with total richness but often contrasts spatially with endemics (Bjarnason et al., 2017). Thus, total richness is a useful but imperfect surrogate, and within-island replication is essential to capture spatial heterogeneity.

*4.1. 2 Topography*

Elevation range was the strongest predictor of endemic richness, underscoring the role of topographic complexity in creating habitat diversity and micro-refugia (Médail & Diadema, 2009; Camilleri et al., 2024). Rugged terrain reduces extinction risk, promotes speciation, and limits disturbance (Irl et al., 2014, 2015; Ye et al., 2021). The consistency of these patterns globally (Irl et al., 2015) suggests topographic complexity is a general driver of island endemism.

*4.1.3 Climate*

Climatic ranges, particularly in precipitation and temperature, predicted endemism more strongly than mean values, indicating adaptation to variability and speciation across fluctuating conditions (Kougioumoutzis et al., 2020; Cutts et al., 2023). Temperature of the coldest quarter was most influential, highlighting frost tolerance as a limiting factor. While endemics are often climate-sensitive (Manes et al., 2021), persistence through past fluctuations (Markonis et al., 2016; Lécuyer et al., 2018; Vicente-Serrano et al., 2025) suggests resilience to variability. At this scale, climate correlates strongly with elevation, emphasizing the need to disentangle collinear predictors (Lawlor et al., 2024). Future work should integrate multiple variables to better anticipate climate change impacts (Moustakas et al., 2025).

*4.1.4 Land Cover and Soil*

Land cover and soil richness contributed to endemic richness but less than topography or climate. Heterogeneity, rather than specific categories, was most important (Thomsen et al., 2022). Although most endemics are in natural habitats, some persist in artificial landscapes (e.g. *Petromarula pinnata* in ruins; Rackham & Moody, 1996), showing adaptive flexibility (McKinney, 2002).

*4.2 Implications for Ecosystem Services*

On Crete, endemic plants are inextricably linked to ecosystem services, reflecting the island's topographic complexity, climatic variability, and long isolation that fostered high levels of endemism (Kougioumoutzis *et al.*, 2020; Stagiopoulou *et al.*, 2025). Many of these taxa (e.g., *Origanum dictamnus*, *Zelkova abelicea*) are of conservation concern under the EU Habitats Directive and simultaneously underpin key provisioning, regulating, cultural, and supporting services. Provisioning services include the production of medicinal herbs, honey, and essential oils, while regulating services are evident in the role of upland endemics in stabilizing soils and buffering microclimatic extremes. Endemics in Crete also sustain cultural values through their integration into folklore, cuisine, and nature-based tourism, while supporting services include habitat provision for specialized pollinators and indicators of habitat quality.

The vulnerability of these services mirrors the threats facing Crete's endemic flora, including overgrazing (Kairis *et al.*, 2015), fires in protected areas (Moustakas, 2025), invasive species (Bjarnason et al. 2017), and climate change (Zittis *et al.*, 2025). Their loss would diminish local livelihoods, cultural identity, and ecological stability. This case study illustrates that the ecological and evolutionary processes driving endemism—particularly topographic and climatic heterogeneity—also underpin ecosystem service provision in islands as well as mainland highlands (Langle-Flores & Quijas, 2020; Sun *et al.*, 2021). Protecting endemic plants is therefore not only essential for conserving evolutionary heritage but also for maintaining the provisioning, regulating, cultural, and supporting services that sustain human well-being (Levis *et al.*, 2024; Luo *et al.*, 2024).

*4.3 Management and policy implications*

Our results have clear implications for biodiversity management and policy in Crete and other Mediterranean islands. Areas of high endemic richness, particularly mountainous and climatically variable regions, should be prioritized for conservation, as they support both unique species and critical ecosystem services (Kougioumoutzis *et al.*, 2025; Stagiopoulou *et al.*, 2025). Management strategies should account for within-island spatial heterogeneity, avoiding reliance on total species richness alone as a surrogate for endemism. Policies must also address emerging threats from tourism, mountainous renewable energy development, and land-use change, integrating habitat protection with sustainable development planning (Kati *et al.*, 2021; Moustakas *et al.*, 2023; Biasi *et al.*, 2024; Leka *et al.*, 2024). The views of island stakeholders need to be accounted for regarding the land use-climate change-biodiversity nexus (Moustakas et al. 2026). Incorporating environmental heterogeneity and climate variability into conservation zoning, alongside monitoring of invasive species, can enhance the resilience of endemic-rich ecosystems (Escobar-Camacho *et al.*, 2021). Finally, cross-island collaboration and the use of transferable predictive models can inform regional conservation strategies beyond Crete, supporting evidence-based decision-making for geologically and biogeographically similar islands (Deo et al., 2024).

*4.4 Uncertainties and future research*

Despite these insights, several limitations remain. Correlations among environmental variables, particularly elevation and climate, may obscure independent effects (Evans et al., 2014; Lawlor et al., 2024), while coarse land cover and soil data may underestimate their influence on endemics. Our focus on species richness excluded functional and phylogenetic diversity, which could reveal additional conservation-relevant patterns. Biotic interactions, such as competition, herbivory, and invasive species, are difficult to quantify at landscape scales but strongly affect species persistence (Bjarnason et al., 2017; Lazarina et al., 2019). Environmental layers may also miss fine-scale microhabitats, refugia, or local anthropogenic impacts, especially in rugged terrain (Irl et al., 2015; Camilleri et al., 2024). Climate-driven projections remain uncertain due to model variability and potential non-linear ecological responses (Moustakas et al., 2025). Future research should combine high-resolution environmental data, multiple biodiversity dimensions, and species-specific responses to better predict endemic persistence under global change.

Although neural networks provided robust explanatory insights, alternative machine learning or ensemble approaches could be explored (Fisher et al., 2024; Sakti et al., 2024). Model sensitivity to training/testing partitioning (Moustakas and Davlias 2021) remains unassessed due to the modest sample size (162 cells). Finally, predictors identified in Crete may not generalize to all islands, though pre-trained models could transfer to geologically and biogeographically similar islands, such as Kasos and Karpathos (Deo et al., 2024).

**5. Conclusions**

  Endemic richness in Crete is strongly influenced by topographic complexity and climatic variability, reflecting distinct ecological and evolutionary processes that are not fully captured by total species richness. Elevation range and climatic heterogeneity create diverse habitats and micro-refugia, promoting speciation and buffering extinction risks, while land cover and soil contribute more weakly, with heterogeneity being more important than specific types. Endemic hotspots only partially overlap with total richness, highlighting the limits of using total richness as a surrogate. These environmentally heterogeneous areas also provide critical ecosystem services, including water regulation, soil stabilization, and cultural value, which are threatened by tourism, renewable energy development, and land-use change. Our findings underscore the need to prioritize mountainous and climatically variable areas in conservation planning, integrating biodiversity, ecosystem services, and predictive modelling to safeguard endemic species and the ecological functions they support under ongoing anthropogenic and climate pressures.